\documentclass[color,french,12pt]{article}
\usepackage{amsmath,amssymb}
\usepackage{fancybox}
\usepackage[latin1]{inputenc}
\usepackage{amsmath,amssymb}
\usepackage{graphicx}
\usepackage{array}
\usepackage{multimedia}
\usepackage[latin1]{inputenc}
\usepackage{euscript}
\usepackage{appendix}
\usepackage{amsthm}
\usepackage{minitoc}
\usepackage{fancyhdr}

\input{tcilatex}
\begin{document}

\title{On Schwarzschild black holes in a D-dimensional noncommutative space}
\author{M. Chabab$^1$ , H.EL Moumni$^1$, M.B. Sedra$^2$ \and {\small 1 :
LPHEA University of Cadi Ayyad Marrakech, FSSM, Department of
physics, Morocco.} \and {\small 2 : LHESIR, Ibn Tofail University at
Kenitra, FSK, Department of physics, Morocco.}}

\maketitle

\begin{abstract}
This work aims to implement the idea of noncommutativity in the subject of
black holes. Its principal contents deal with a study of Schwarzschild black
holes in a D-dimensional noncommutative space. Various aspects related to
the non commutative extension are discussed and some non trivial results are
derived.
\end{abstract}

\section*{Introduction}

String theory $\cite{1,2,3}$, noncommutative geometry
$\cite{4,5,6,7,8}$ as well as Black holes $\cite{9,10, AS}$ were
extensively studied recently in different contexts. Along these
lines of research, we consider that the adaptation of
noncommutativity in the subject of black holes, as discussed also by
several previous works $\cite{11,12,13,14,15,16}$, is not only a
simple technical exercise. Noncommutativity is a mathematical and
physical concept that can be variously motivated. Perhaps the
simplest is that it might improve the renormalizability properties
of a theory at short distances or even render it finite. Its also
considered as fundamental in expressing uncertainty in quantum
mechanics. Another motivation is the belief that in quantum theories
with gravity, space-time must change its nature at the Planck scale
distance. Quantum gravity has an uncertainty principle which
prevents one from measuring positions to better accuracies than the
Planck length: the momentum and energy required to make such a
measurement will itself modify the geometry at these scales
$\cite{2}$.

The simplest noncommutativity one can postulate is the commutation relations
\begin{equation}
\lbrack \hat{x}_{\mu },\hat{x}_{\nu }]=i\theta _{\mu \nu },  \label{1}
\end{equation}
where $\hat{x}_{\mu }$ are the space-time coordinates operators and $%
\theta_{\mu \nu }$ is an antisymmetric constant tensor of dimension $%
(length)^{2}$. In this context, the usual product of fields should be
replaced by the star-product defined by:
\begin{equation}
\left( f\star g\right) =\exp \left( \frac{i}{2}\theta _{\mu \nu }\frac{%
\partial }{\partial x^{\mu }}\frac{\partial }{\partial y^{\nu }}\right)
f(x)g(y)|_{x=y},  \label{2}
\end{equation}

where $f$ and $g$ are two arbitrary functions assumed to be infinitely
differentiable. By virtue of this definition, we can write:
\begin{equation}
\left[ {\hat{x}}_{i},{\hat{x}}_{j}\right] =i\theta _{ij},\;\;\;\left[ {\hat{x%
}}_{i},{\hat{p}}_{j}\right] =i\delta _{ij},\;\;\;\left[ {\hat{p}}_{i},{\hat{p%
}}_{j}\right] =0.  \label{3}
\end{equation}
We focus then in this work to study Schwarzschild black holes in a $D$%
-dimensional noncommutative space. Various aspects related to the non
commutative extension are discussed and some non trivial results are derived.

\section*{Schwarzschild black holes in a D-dimensional noncommutative space}

Consider the metric of Schwarzschild black hole in $D$ dimensions $\cite{15}$%
.
\begin{equation}
ds^{2}=-\left( 1-\left( \frac{r_{H}}{r}\right) ^{D-3}\right) dt^{2}+\frac{%
dr^{2}}{1-\left( \frac{r_{H}}{r}\right) ^{D-3}}+r^{2}d\Omega _{D-2}^{2}
\label{4}
\end{equation}%
where
\begin{equation}
r_{H}^{D-3}=\frac{16\pi MG}{(D-2)\Omega _{D-2}},\;\;\Omega _{D}=\frac{2\pi
^{(D+2)/2}}{\Gamma \left( \frac{D+1}{2}\right) }
\end{equation}%
where for $D=4,$ $\ r_{H}=2MG$ is nothing but the radius of the horizon with
the use of the conventions $\hbar =c=k_{B}=\frac{1}{4\pi \epsilon _{0}}=1$
and $G=m_{Pl}^{-2}=\ell _{Pl}^{-2}$ where $m_{P_{l}}$ and $\ell _{P_{l}}$
are the Planck mass and the Planck length, respectively. Next, for the
noncommutative version of the $D$-Schwarzschild black hole's metric, we
propose the following generalized form
\begin{equation}
ds^{2}=-\left( 1-\left( \frac{r_{H}}{\sqrt{\hat{r}\hat{r}}}\right)
^{D-3}\right) dt^{2}+\frac{d\hat{r}d\hat{r}}{\left( 1-\left( \frac{r_{H}}{%
\sqrt{\hat{r}\hat{r}}}\right) ^{D-3}\right) }+\hat{r}\hat{r}d\Omega
_{D-2}^{2}  \label{8}
\end{equation}%
where $\hat{r}$ has to satisfy the constraint
\begin{equation}
1-\left( \frac{r_{H}}{\sqrt{\hat{r}\hat{r}}}\right) ^{D-3}=0  \label{9}
\end{equation}%
with respect to $(\ref{3})$. These considerations lead to a new $(x,p)$%
-coordinate system,
\begin{equation}
x_{i}={\hat{x}}_{i}+\frac{1}{2}\theta _{ij}{\hat{p}}_{j},\;\;\;p_{i}={\hat{p}%
}_{i},  \label{10}
\end{equation}%
where the new variables are shown to satisfy, once again, the usual
canonical commutation relations:
\begin{equation}
\left[ x_{i},x_{j}\right] =0,\;\;\;\left[ x_{i},p_{j}\right] =i\delta
_{ij},\;\;\;\left[ p_{i},p_{j}\right] =0.  \label{11}
\end{equation}%
Insertion of (\ref{10}) into the relation $(\ref{9})$ yields
\begin{equation}
1-\left( \frac{r_{H}}{\sqrt{(x_{i}-\theta _{ij}p_{j}/2)(x_{i}-\theta
_{ik}p_{k}/2)}}\right) ^{(D-3)}=0  \label{12}
\end{equation}%
and
\begin{equation}
1-\left( \frac{r_{H}}{r}\right) ^{D-3}\left[ 1+\left( \frac{D-3}{2}\right)
\frac{x_{i}\theta _{ij}p_{j}}{r^{3}}+\left( \frac{D-3}{8}\right) \frac{%
\theta _{ij}\theta _{ik}p_{j}p_{k}}{r^{2}}\right] +{\mathcal{O}}(\theta
^{3})+\ldots =0  \label{13}
\end{equation}%
or equivalently
\begin{equation}
1-\left( \frac{r_{H}}{r}\right) ^{D-3}-\frac{r_{H}^{D-3}}{2r^{D-1}}\left[
\left( \frac{D-3}{2}\right) \vec{L}.\vec{\theta}-\left( \frac{D-3}{8}\right)
\left( p^{2}\theta ^{2}-(\vec{p}.\vec{\theta})^{2}\right) \right] +{\mathcal{%
O}}(\theta ^{3})+\ldots =0  \label{15}
\end{equation}

where $\theta _{ij}=\frac{1}{2}\epsilon _{ijk}\theta _{k}$.

As a particular case, setting $\theta =\theta _{\xi }$ and the remaining $%
\theta $-components equal to zero, we can write $\vec{L}.\vec{\theta}=L_{\xi
}\theta $, and then we have

\begin{equation}
r^{D-1}-r_{H}^{D-3}r^{2}-\frac{r_{H}^{D-3}}{4}\left[ (D-3)L_{\xi }\theta
-\left( \frac{D-3}{8}\right) (p^{2}-p_{\xi }^{2})\theta ^{2}\right] +{%
\mathcal{O}}(\theta ^{3})+\ldots =0  \label{16}
\end{equation}%
Using $p^{2}=\sum_{i=1}^{D-1}p_{i}^{2}$ we have $(p^{2}-p_{\xi }^{2})\theta
^{2}=(\sum_{i\neq \xi }^{D-1}p_{i}^{2})\theta ^{2}$, the Schwarzschild black
hole is non rotating, then $\vec{L}=\vec{0}$ implying $L_{\xi }=0$. Thus $(%
\ref{16})$ simplifies to

\begin{equation}
r^{D-1}-r_{H}^{D-3}r^{2}+\frac{r_{H}^{D-3}}{32}(D-3)(\sum_{i\neq \xi
}^{D-1}p_{i}^{2})\theta ^{2}+{\mathcal{O}}(\theta ^{3})+\ldots =0  \label{17}
\end{equation}%
The obtained equation is a $(D-1)$-polynomial equation of type
\begin{equation}
r^{D-1}+a.r^{2}+b=0,  \label{poly}
\end{equation}%
with

\begin{eqnarray}
a &=&-r_{H}^{D-3}  \label{18} \\
b &=&\frac{r_{H}^{D-3}}{32}(D-3)(\sum_{i\neq \xi }^{D-1}p_{i}^{2})\theta ^{2}
\end{eqnarray}

Note that solving this equation is an important task since the obtained
solutions correspond to the horizon radius in $D$-dimensional noncommutative
space. Performing algebraic straightforward computations, we are able to
derive explicit solutions up to $D=5$.

For $D=4,$ we recover the results presented in previous works, see for
instance $\cite{15}$ and references therein.

The case $D=5$ is important since it gives us original results. Indeed,
working the equation $(\ref{poly})$ we can derive two kind of physical
solutions, namely
\begin{equation}
\hat{r}_{h\pm }=\frac{r_{H}^{{}}}{\sqrt{2}}\sqrt{1\pm \sqrt{1-\frac{1}{4}%
\left( \frac{\theta }{r_{H}^{{}}}\right) ^{2}\sum_{i\neq \xi }^{D-1}p_{i}^{2}%
}}  \label{20}
\end{equation}

Herewith some remarks:

- The key point concerns the standard limit which gives $\hat{r}%
_{h_{+}}=r_{H}^{{}}$ once the deformation parameter $\theta $ is set equal
to zero.

- For higher dimensions $\left( D>5\right) $, there is no technical way, to
our knowledge, to solve analytically the above equation $(\ref{poly})$.
However, one can eventually use numerical approaches to approximate
solutions.

- Setting $D=3,$ this case is remarkable and deserves a particular interest.
In fact, one easily observe, from the previous derived expressions $(\ref{18}%
)$, a screening effect on the $\theta -$noncommutativity whenever the choice
of the dimension is taken to be  $D=3$.

- In the same way, we underline that noncommutative charged black
holes with charge $q$ and mass $\mu $ are governed by the following
polynomial equation
\begin{equation}
r^{2(D-2)}+\alpha r^{D-1}+\beta ^{D-3}+\gamma r^{2}+\delta =0,
\end{equation}%
where $\alpha$, $\beta$, $\gamma$ and $\delta$ are constants
depending on the charge $q$, the mass $\mu$ of the Black hole and on
the noncommutativity $\theta$ parameter.

Solving this equation will be discussed explicitly in our
forthcoming work.

\end{document}